\documentclass[letterpaper, 10 pt, conference]{ieeeconf}  

\IEEEoverridecommandlockouts                              

\usepackage{graphics} 
\usepackage{epsfig} 
\usepackage{mathptmx} 
\usepackage{times} 
\usepackage{amsmath} 
\usepackage{amssymb}  

\usepackage{color}

\newcommand{\newText}[1]{{\color{black}{#1}}}
\newcommand{\camready}[1]{{\color{black}{#1}}}

\title{\LARGE \bf
\newText{Investigating Social Haptic Illusions for Tactile Stroking (SHIFTS)} 
}

\author{Cara M. Nunez$^{1,2}$, Bryce N. Huerta$^{1}$, Allison M. Okamura$^{1}$ and Heather Culbertson$^{3}$
\thanks{*This work was supported in part by the National Science Foundation Graduate Fellowship Program and National Science Foundation grant 1830163.}
\thanks{$^{1}$Department of Mechanical Engineering and $^{2}$Department of Bioengineering, Stanford University, Stanford, CA 94305.}
\thanks{$^{3}$Department of Computer Science, University of Southern California, Los Angeles, CA, 90089.}
\thanks{(email: \{nunezc; bhuerta; aokamura\}@stanford.edu, hculbert@usc.edu).}
}

\begin{document}

\onecolumn
\textcopyright~2020 IEEE.  Personal use of this material is permitted.  Permission from IEEE must be obtained for all other uses, in any current or future media, including reprinting/republishing this material for advertising or promotional purposes, creating new collective works, for resale or redistribution to servers or lists, or reuse of any copyrighted component of this work in other works.
\newpage
\twocolumn

\maketitle
\thispagestyle{empty}
\pagestyle{empty}

\begin{abstract}

A common and effective form of social touch is stroking on the forearm. We seek to replicate this stroking sensation using haptic illusions. This work compares two methods that provide sequential discrete stimulation: sequential normal indentation and sequential lateral skin-slip using discrete actuators. Our goals are to understand which form of stimulation more effectively creates a continuous stroking sensation, and how many discrete contact points are needed. We performed a study with 20 participants in which they rated sensations from the haptic devices on continuity and pleasantness. We found that lateral skin-slip created a more continuous sensation, and decreasing the number of contact points decreased the continuity. These results inform the design of future wearable haptic devices and the creation of haptic signals for effective social communication.

\end{abstract}

\section{INTRODUCTION}

Touch is a critical aspect of interpersonal communication\newText{, especially the communication of emotion between humans}
~\cite{app2011nonverbal}. It is a crucial component of daily life and is essential to human development, communication, and survival~\cite{Chang2001TouchInCaring,Field2002TouchInfants,Herenstein2002TouchInfancy,Weiss2004MaternalTactileStim,Cullen2002TouchAutism,Dunbar2010SocialRoleOfTouch,Field2010EmotionalPhysicalWellBeing,Gallace2008InterpersonalTouch,Henricson2008TouchIntensiveCare,Whitcher1979TherapeuticTouch}. A major challenge in the field of haptics is how to provide meaningful and realistic sensations\newText{, which are currently lacking in most computer-mediated interactions~\cite{vanerp2015social},} that are similar to what is relayed during social touch interactions. 
The inability to transmit touch during interpersonal communication leads to a limited feeling of social presence during virtual interactions between people, motivating the design of haptic systems to deliver virtual social touch cues. This requires an understanding of the characteristics of social touch, leading to the design and selection of control parameters for haptic systems to emulate social touch. 
\newText{With a stronger understanding of the parameters involved in imitating human touch, we can develop wearable haptic devices for mediated social touch~\cite{haans2006mediated} or communication between a human and robot~\cite{huisman2017social}.}

Specific types of tactile stimulus, such as stroking, squeezing, and tapping, convey different emotional messages like love, happiness, or gratitude, and can be used by humans to successfully identify the message being relayed~\cite{hertenstein2009communication,hertenstein2006touch,thompson2011effect}. The tactile stimulus in many of these cases comes from the contact of the toucher's hand with the receiver's forearm~\cite{hertenstein2006touch,Hauser2019WHC,Suvilehto2015TopographyOfTouch}. 
\newText{Mechanoreptors within the skin respond to the different stimuli, such as vibrations, skin deformation, and skin stretch~\cite{johnson2001roles}, that are involved in social touch. Researchers have shown that one mechanoreceptor in particular, the C tactile (CT) afferents, exists in the forearm and is involved in emotional touch~\cite{Olausson2002}. The CT afferents respond optimally to light, stroking sensations at slow speeds in the range of 1-10~cm/s~\cite{mcglone2014discriminative}. Ackerley et al.\ also showed that stroking sensations on the forearm with speeds of 1-10~cm/s were rated to be more pleasant than slower or faster speeds~\cite{ackerley2014touch}. }
In our investigation, we aim to create a continuous, pleasant stroking sensation similar to that of human touch by activating 
a combination of mechanoreceptors via the use of \newText{social haptic illusions for tactile stroking (SHIFTS).}

This paper begins with a discussion of prior work on the development of haptic devices for social touch and the challenges and benefits of wearable haptic devices for this application. We then describe two SHIFTS devices that use haptic illusions involving sequential discrete stimulation to create a continuous, pleasant stroking sensation. Next, we conduct a user study to directly compare the performance of the two devices. In our experiment, we also investigate how the perception of the sensation changes when the number of contact points is reduced. We conclude by discussing how the results of our study impacts the development of future wearable haptic devices for social communication.

\section{RELATED WORK}
\label{sec:relatedwork}
Here we discuss previous haptic devices designed for social touch. We also give an overview of existing haptic solutions for creating stroking sensations and their inherent limitations. We then review guidelines for designing wearable haptic devices. Finally, we describe prior uses of haptic illusions and highlight their potential for creating continuous stroking sensations involved in social touch.

\subsection{Haptic Devices for Social Touch}

\newText{Researchers have designed haptic devices to recreate specific social interactions, such as a hug~\cite{Cha2009HugMe,Delazio2018Pneumatic,mueller2005hug,tsetserukou2010haptihug,Yu2015PARO} or handshake~\cite{nakanishi2014remote}. }
Other haptic devices attempt to directly replicate an input signal~\cite{brave1997intouch} or map between different modalities, such as force input to vibration output~\cite{huisman2013tasst}, to create various forms of social touch. Still other haptic displays use several miniature robots to coordinate movement to create varying social touch sensations~\cite{Dementyev2016BodyRobots,Kim2019SwarmHaptics}. 

\newText{Several social haptic devices have been designed to create a stroking sensation using a variety of actuation techniques. Researchers have explored directly stimulating the skin using lateral motion~\cite{eichhorn2008stroking,knoop2015tickler,Moriyama2018Fingers}. }
However, stroke length is limited in these direct stimulation devices (1~mm to 1~cm). \newText{Creating a long stroking sensation using direct lateral stimulation would require more complex actuation and mechanical design which would likely result in a set-up that is heavy, bulky, and would be difficult to implement in a wearable device. }
\newText{As an alternative, one research group created a stroking sensation using an air jet~\cite{tsalamlal2015haptic}. }
Similarly, this is difficult to implement into a wearable device because it requires access to compressed air. These limitations have led researchers to create the illusion of motion across the skin using vibration~\cite{israr2011tactile,Raisamo2013Comparison}, \newText{and investigate its potential as a social haptic device in creating a stroking sensation~\cite{huisman2016simulating,Israr2018VibGrid,seifi2013first}. }
However, vibrations alone do not realistically display the signals used in social touch.

\subsection{Wearable Haptic Devices and Haptic Illusions}
\newText{Wearable haptic devices make it possible for haptic feedback to be provided in different locations in space or while the user is moving, instead of requiring the user to remain in a specific location. }
Additionally, wearable haptic devices enable unobtrusive and private communication. \newText{However, when designing wearable devices one must not think only about desired technical features of the device (such as force output), but also functional aspects of the device (such as weight and comfort)~\cite{Pacchierotti2017wearablereview}. }
\newText{Since the actuators are usually the most bulky and heavy components of a haptic device, actuator selection is an important part of wearable device design and requires the designer to make decisions regarding trade-offs between force, resolution, and workspace among many others. }
\newText{We believe designers can bypass these trade-offs with haptic illusions, which use small and lightweight actuators to create sensations that would normally require actuators that are impractical for a wearable device.}

The most well-known haptic illusion is sensory saltation, or ``
the cutaneous rabbit" effect~\cite{Geldard1972sensorysaltation}. Researchers used discrete vibration to create the effect of a rabbit hopping along the forearm. This leverages the sparse distribution of mechanoreceptors on the forearm and tricks one into thinking that \newText{a }
rabbit is hopping along the skin. The use of vibration to create the illusion of motion across the skin for social touch applications follows a similar principle. Previously, we investigated the use of sequential discrete normal indentation~\cite{VCstudy} and sequential discrete lateral skin-slip~\cite{NunezToH2019} to create the illusion of tactile stroking. Another group developed a wearable haptic sleeve that uses pneumatic actuators to provide sequential normal indentation~\cite{Wu2019PneumaticSleeve} and compared its performance to the device developed in~\cite{VCstudy}. The success of these haptic illusions led another group to develop a multi-dimensional tactile display that can relay discrete vibration, pressure, and shear stimuli and investigate the effect of the combination of these actuation techniques on the illusion of tactile stroking~\cite{Kim2019VPS}. While these haptic illusions effectively show that it is possible to create tactile stroking, they do not investigate how many contact points are necessary in order to create this sensation or look at whether fewer contact points are needed depending\camready{ on} 
the type of stimulus. Answering this question would provide more information about the sensation that is created and could also further reduce the components needed within the device, thereby reducing weight and power consumption.

\section{SHIFTS DEVICES}
\label{sec:design}

\begin{figure}[b]
\centering
  \includegraphics[width=\columnwidth]{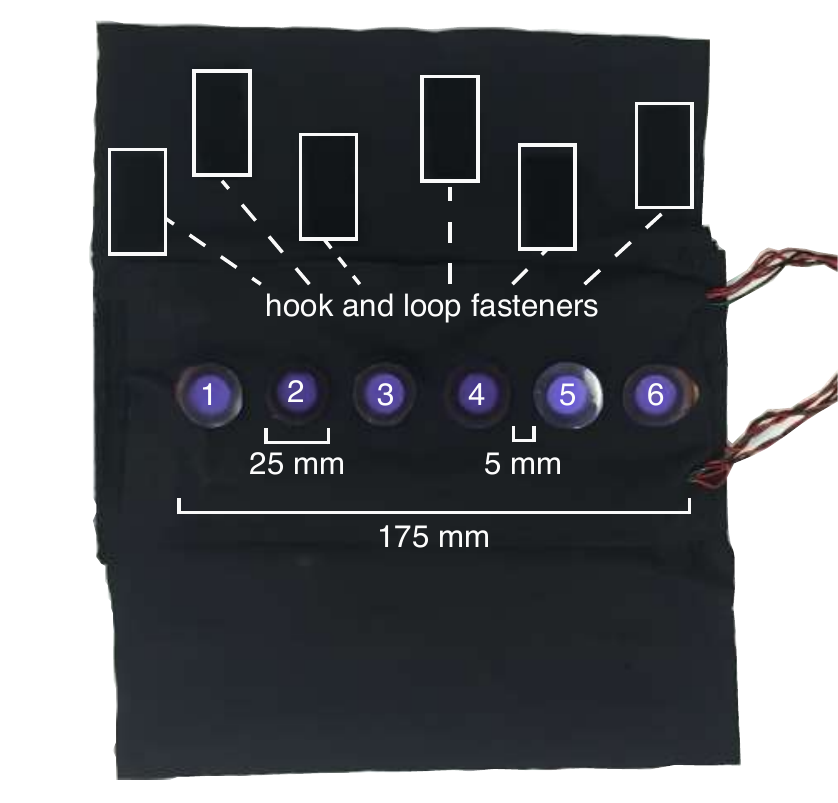}
  \caption{Voice coil SHIFTS device. We have identified the dimensions of the contact area, space between contact points, and total stroke length in addition to marking the location of the hook and loop fasteners.}~\label{fig:voicecoilDevice}
\end{figure}

\begin{figure}[b!]
  \centering
  \includegraphics[width=0.9\columnwidth]{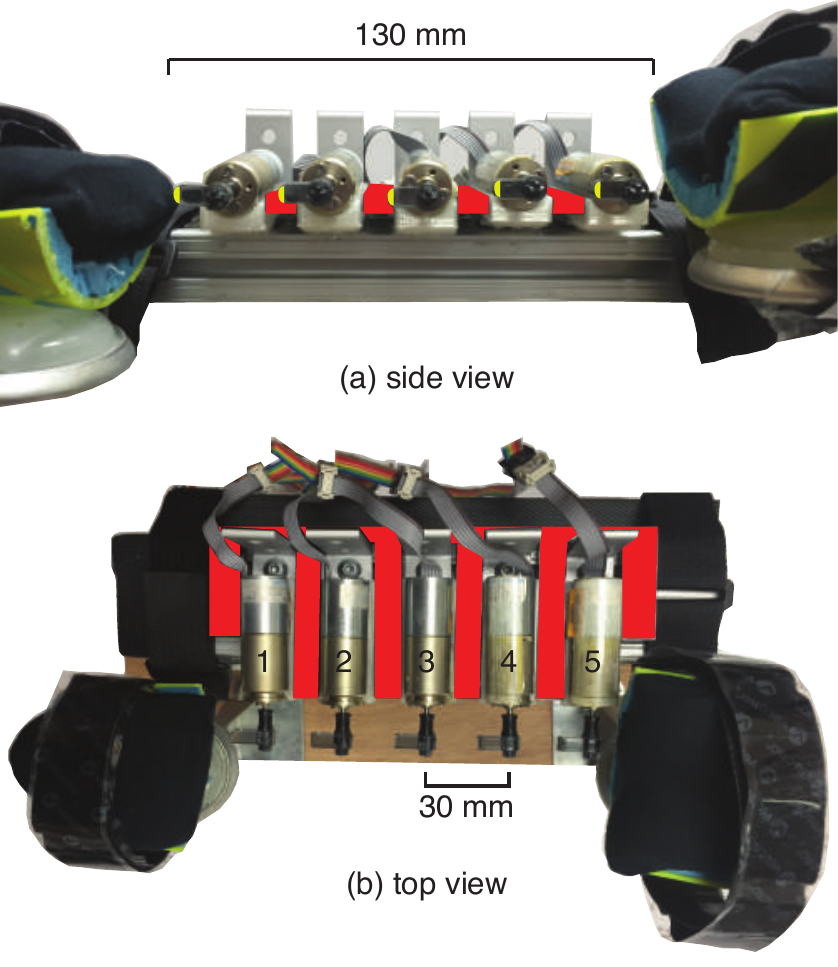}
  \caption{Motor SHIFTS device. (a) A side view shows the starting position of the tactors and the marks on the tactors (in yellow) used to align the participants for consistent indentation of 1.5~mm. \newText{We have also identified the total stroke length, 130~mm, when all 5 motors are actuated.} (b) A top view identifie\camready{s }
  the consistent spacing between the motors, 30~mm, via the acrylic separator (in red).}~\label{fig:motorDevice}
\end{figure}

This section describes the design of the two SHIFTS devices, which can be seen in Figure~\ref{fig:voicecoilDevice} and Figure~\ref{fig:motorDevice}. The SHIFTS devices were previously developed and tested to determine some of the parameters involved in creating a continuous and pleasant stroking sensation involved in social touch~\cite{VCstudy,NunezToH2019}. The original ideas for these devices came from simple hand-actuated prototypes developed following the principles of haptic sketching~\cite{moussette2012simple}. They were then constructed as electromechanical prototypes, mimicking the actuation of the haptic sketches, that could be programmed and controlled. In this section, we will summarize the device designs, key results from the previous investigations, and parameters of the devices that will be held constant during the user study we conducted in this paper.

\subsection{Voice Coil Device}
The voice coil SHIFTS device~\cite{VCstudy} consists of six voice coil actuators (Tectonic Elements TEAX19C01-8) arranged in a 1-D linear array, as shown in Figure~\ref{fig:voicecoilDevice}. \newText{While voice coils are usually actuated at high frequencies to create vibrations, the voice coils in this array are actuated at low frequencies ($<5$~Hz) to create normal indentation. }
\newText{Following the device design in~\cite{VCstudy}, the voice coil array is embedded into an elastic sleeve that is comfortable for the user and can be adjusted to fit the forearm of different users. }
We sewed in a layer of inelastic, but flexible, canvas to the portion of the sleeve directly behind the actuators \newText{to provide a stronger backing in order to ensure that the force of the voice coil actuators are directed into the user's skin}. 
We sewed hook and loop fasteners into the sleeve so that we could wrap the sleeve around the user's forearm and \newText{tighten it.}

\newText{When purchased off-the-shelf, the contact area of the voice coils is a thin ring. Similar to~\cite{VCstudy}, we added a circular piece of thin polypropylene to the ring in order to distribute the force consistently across the contact area with the skin. }
The diameter of each contact point is 25~mm. The voice coils are placed directly next to each other in the array, creating 5~mm of space between each contact point. This arrangement of the voice coils allows for a stroke length of 175~mm (6.89~in) when all six voice coils are actuated.

\newText{To use the voice coil SHIFTS device, the sleeve containing the voice coil actuators is wrapped around the user's forearm and tightened using the hook and loop fasteners to prevent slipping or movement of the device. Due to this attachment method, the voice coils begin in contact with the skin. }
To create the haptic illusion of a stroking sensation, \newText{the actuators must move backwards off of the skin before }
applying normal indentation. As discussed in detail in~\cite{VCstudy}, the \newText{actuators }
retract and then indent into the skin following a quadratic profile. \newText{Each voice coil is controlled by an analog output pin from a Sensoray board connected to a Sensoray 826 PCI card (updated at 1000~Hz). The signals from the analog output pins are each passed through linear current amplifiers specially-made with power op-amps (LM675T) to implement a gain of 1~A/V. }
This form of actuation provides at least 1.5~mm depth of skin indentation which was shown in~\cite{Biggs2002resolution} to be consistently and accurately perceived by a user.

\newText{The voice coil SHIFTS device sequentially indents the actuators into the forearm to create the sensation of a stroke along the arm. }
\newText{The stroke sensation can be controlled by varying the indentation duration and the amount of delay between the onset of indentation for adjacent actuators~\cite{VCstudy}. }
\newText{In this previous study, }
we investigated different combinations of indentation duration and amount of delay to assess their effect on user perception of a continuous and pleasant stroking sensation on the dorsal forearm and dorsal upper arm. The sensation was rated as more continuous on the forearm than the upper arm, and perceived continuity increased with increasing indentation duration and decreased with increasing delay. Pertaining to pleasantness, there was no significant difference between the forearm and the upper arm, and the  pleasantness of the interaction was highest for shorter delays and increased as the indentation duration increased. From the investigation, we found that the highest continuity and pleasantness ratings both occurred on the dorsal forearm with an indentation duration of 800~ms and a 12.5\% delay between onset of indentation of adjacent actuators. This combination corresponds to an effective speed of travel up the arm of 13.5~cm/sec, which falls slightly above the optimal range for activating the CT afferents~\cite{mcglone2014discriminative,ackerley2014touch}.\camready{ However, this speed corresponds to the stroking velocity humans use when instructed to stroke the forearm in a way that a person would like it~\cite{Croy2015Stroking}.} 
Thus, for our investigation in this paper we held these conditions constant so that we could investigate how changes in number of actuators affect the best-performing condition.

\subsection{Motor Device}
The motor SHIFTS device~\cite{NunezToH2019} consists of five rotary motors (Faulhaber 1624E0175 DC motors with a quadrature encoder) arranged in a 1-D linear array, as shown in Figure~\ref{fig:motorDevice}. \newText{Rounded tactors were attached to the motor shafts and are the component that makes contact with the user to provide skin-slip. }
The tactors were laser-cut from 1/4-inch acrylic and \newText{each adhered to a coupler with a + shaped cross-section. The couplers were then press-fit onto the shafts of the motors. The couplers helped to prevent the tactors from rotating in response to the torque generated from contact with the skin. }

Unlike the voice coil SHIFTS device, the motor SHIFTS device is not wearable. \newText{Following the design in~\cite{NunezToH2019}, the motors are secured in 3-D printed holders to anchor the motors in place and keep the round motors from shifting. The motors are located between two adjustable stands, one for the wrist and one for the elbow, which hold the forearm in place and allow the height to be adjusted for each user to ensure each tactor will indent 1.5~mm into the skin~\cite{Biggs2002resolution}. }

\newText{The motor SHIFTS device sequentially rotates the motors such that the tactors provide discrete skin-slip sensations~\cite{NunezToH2019}. Each tactor starts off of the skin (as shown in Fig.~\ref{fig:motorDevice}) and rotates to make contact with the skin. The tactors initially make normal contact with the skin and then slide along the skin as they continue rotating until they slip off of the skin to create the short skin-slip sensation. The tactors continue rotating back to their starting position off of the skin. }
\newText{These short skin-slip sensations combine to create a longer stroking sensation. }
The encoder values from each motor are passed through a PID controller to set the position of the tactor. Each motor is driven by a separate analog output from a Sensoray 826 PCI card which is updated at 10~kHz. Similar to the voice coil SHIFTS device, we pass the analog output through a custom-built linear current amplifier using a power op-amp (LM675T) with a gain of 1~A/V.

The array of motors creates the sensation of a stroke along the forearm by sequentially indenting the actuators into the arm and providing lateral skin-slip. The feeling of this stroke can be controlled by varying the rotation speed of the tactors (angular velocity) and the amount of delay between the onset of rotation for adjacent tactors. In a previous study~\cite{NunezToH2019}, we investigated the combination of the angular velocity and delay to determine the effect on user perception of a continuous and pleasant stroking sensation on the dorsal and volar forearm. We found that the perceived continuity did not differ between the dorsal and volar forearm, but the sensation was more pleasant when applied to the volar forearm. We further determined that the sensation was perceived most continuous and pleasant when using an angular velocity of 0.66$\pi$~rad/sec and a delay of 10\%. Thus, for our investigation in this paper we held these conditions constant and applied the sensation to the volar forearm so that we could investigate how device changes affect the best performing condition.

As a follow-up to the initial study, we investigated the effect that distance between contact points had on creating the haptic illusion of tactile stroking~\cite{NunezToH2019}. While we hypothesized that increasing the distance between the contact points would negatively affect the illusion of a continuous and pleasant stroking sensation, we found that there was actually no significant difference in the perceived sensation. Because the motors had a diameter of 20~mm, this was the distance used when assessing the actuation parameters in the first study. To increase the distance between contact points, we needed to reduce the number of contact points from five to four so the sensation could still fit in the workspace of the forearm. Since we found that slightly reducing the number of contact points and increasing the spacing between contact points did not affect the perceived continuity or pleasantness of the sensation, we were curious about the minimum number of contact points that are necessary to effectively create tactile stroking. To investigate this question, we kept the distance between contact points at 30~mm in the device we study in this paper. We laser-cut a piece of acrylic which slides between the motor carriages to ensure consistent spacing. We used the tactor design discussed in detail in~\cite{NunezToH2019}, which results in each tactor traveling 10~mm along the skin. When all five motors are actuated, this results in a stroke length of 130~mm (5.12~in).

\section{USER STUDY}
\label{sec:study}
We conducted a study to better understand the type of stimulation that effectively creates a continuous stroking sensation. We applied haptic sensations to participants via the SHIFTS devices described in \newText{Section~\ref{sec:design} }
using haptic illusions to create a continuous and pleasant stroking sensation. In addition 
to directly \newText{comparing }
these two actuation techniques, we also investigate how many contact points are necessary to create the illusions.

\subsection{Hypotheses}
We hypothesize that the motor device, which provides sequential discrete lateral skin-slip, will result in a more continuous sensation because of the inclusion of direct lateral motion, as compared to the normal indentation applied by the voice coil device. We also hypothesize that reducing the number of contact points would reduce the perceived continuity and that a minimum number of contact points is necessary to maintain the illusion. Previous research~\cite{NunezToH2019} showed that the continuity and pleasantness of the stroking sensation was maintained even when increasing the spacing between the contact points and decreasing the number of contact points. Based on these previous results, we hypothesize that increasing the number of contact points will increase the perceived continuity of the sensation, but the continuity will eventually plateau.

\begin{figure}[b]
\centering
  \includegraphics[width=\columnwidth]{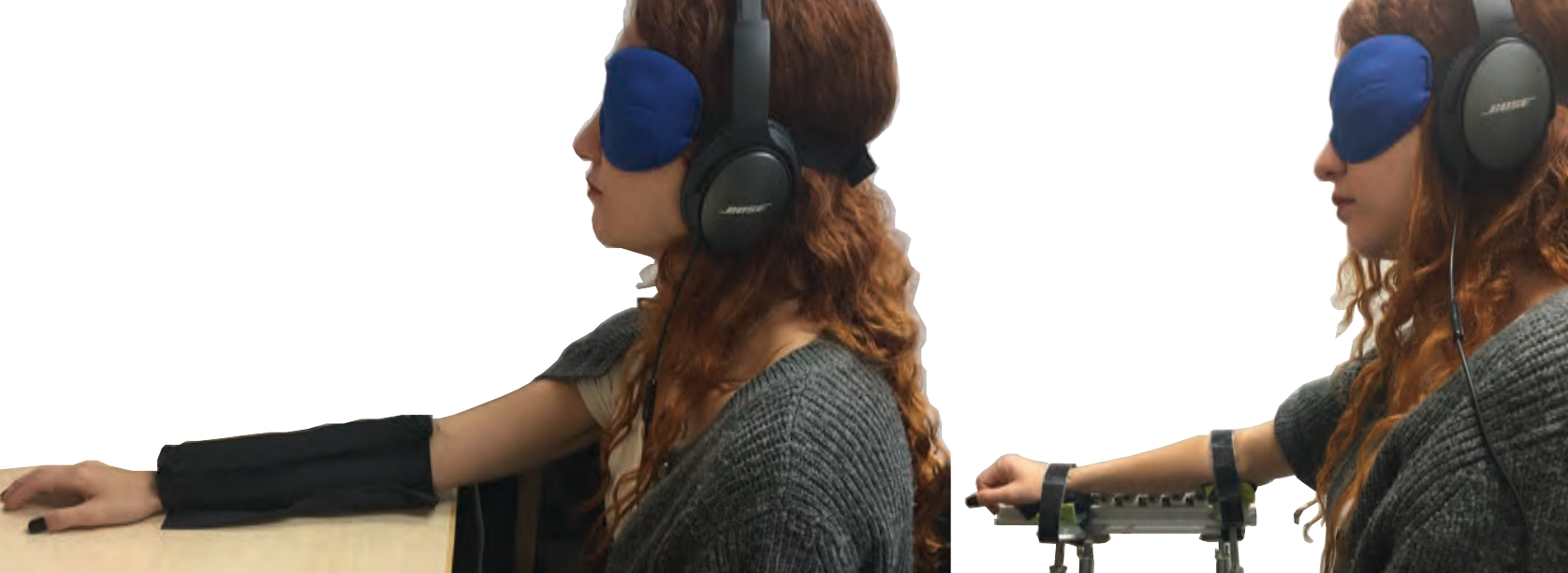}
  \caption{Experimental setup of the user study showing the participant's arm placed in the voice coil SHIFTS device such that they feel the stimuli on the dorsal forearm and rest their arm directly on the table in front of them (left) and placed in the motor SHIFTS device such that they feel stimuli on the volar forearm and have their arm placed slightly off to the side for comfort (right). Participants wore a blindfold and noise-cancelling headphones playing white noise.}
  \label{fig:experimentSetup}
\end{figure}

\subsection{Experimental Setup}
The experimental setup consisted of a desktop computer with a Sensoray 826 PCI card used to control both haptic devices (the voice coil and motor devices) and their corresponding drive circuitry. Figure~\ref{fig:experimentSetup} shows how the participant interact\camready{s} with the haptic devices while seated. Participants would wear the voice coil haptic device on their right arm such that the voice coils would contact their dorsal forearm; they would then rest their forearm on the table in front of them. The tightness of the voice coil sleeve device was adjusted such that the voice coils were securely in place on the participant's forearm and did not shift when the participant picked up their arm. For the motor haptic device, participants would place their right arm on top of the wrist and elbow rests to feel the haptic stimuli applied to their volar forearm. \newText{We adjusted the height of the elbow and wrist of each participant to ensure that there was consistent indentation into the skin for each tactor, before strapping the participant into the device with hook and loop fasteners.} 
The end effector of each tactor was marked at a depth of 1.5~mm, allowing us to check that the tactor would indent 1.5~mm by manually rotating each tactor \newText{such that it was perpendicular and indented into the participant's skin.} This helped to ensure that the indentation profile was consistent across tactors and across participants. \newText{After checking to ensure that each tactor would indent 1.5~mm, we manually rotated the tactors to their starting position off of the skin (shown in Fig.~\ref{fig:motorDevice}).} 

The participants were not allowed to see the devices before or during the experiment. The devices were hidden under a black sheet until the participants were ready to begin the study and had put on a blindfold. During the study, they also wore Bose QuietComfort 25 noise cancelling headphones playing white noise to prevent auditory distractions or cues.

\subsection{Participants}

Twenty-four participants (8 male, 16 female; aged 18-37) were recruited. All participants were right-hand dominant. None had neurological disorders, injuries to the hand or arm, or other condition\camready{s} that may have affected their performance in this experiment. Participants' previous haptic experience ranged from none to extensive. However, none of the participants had any previous experience with either of the two haptic devices used in the study. They were compensated with a \$15 gift card for their time (approximately 45 min) and the study was approved by Stanford University's Institutional Review Board. Participants provided informed consent.

\subsection{Procedure}

Before the study, we informed the participants that they would experience various touch stimuli on their forearm from haptic devices and would be asked to rate the sensation. Participants completed the study in two phases: one phase using the voice coil device providing sequential discrete normal indentation, and the other phase using the motor device providing sequential discrete skin-slip. The order of the two phases was randomly determined for each participant, and the order was balanced across all participants.

In the study, we varied the number of contact points used to apply the sequential discrete actuation. The voice coil device consisted of 6 contact points, creating 6 possible actuation conditions. Each of these were \newText{randomly} repeated 5 times for a total number of 30 trials. Similarly, the motor device consisted of 5 contact points, creating 5 possible actuation conditions for a total number of 25 \newText{randomized} trials. For both devices, the actuation always began at the wrist. \newText{Participants completed all trials corresponding to one device followed by all of the trials corresponding to the second device.} 
Between the phases, participants were given a 2 minute break and allowed to remove the blindfold and headphones. \newText{Participants finished both phases of the study (55 total trials) in approximately 30 minutes.} 

After feeling the haptic stimulus for each trial, \newText{the participants were asked to rate the sensation on two scales. First, participants rated the perceived continuity of the sensation on a Likert scale from 1 to 7 (where 1 is discrete and 7 is continuous). Second, participants rated the perceived pleasantness of the sensation on a Likert scale from -7 to 7 (where -7 is very unpleasant, 0 is neutral, and 7 is very pleasant).} 
Participants were asked to state their ratings verbally out loud, and their ratings were recorded by the investigator. \newText{Once participants had completed all trials with both devices, they filled out a post-study survey which asked them }
to select which device they preferred, 
provide a written description if they noticed any differences in the sensations between trials and to describe what those differences were, and provide any additional comments.

\subsection{Analysis}

In our experiment, we had two independent variables (type of actuation and number of contact points) and two dependent variables (continuity rating and pleasantness rating). First, to compare the two SHIFTS devices at their original design (the voice coil device with 6 contact points and the motor device with 5 contact points), a Mauchly's Test of Sphericity and one-way repeated measures ANOVA was performed for each dependent variable. Then, to examine the effects of the two independent variables including interaction, a Mauchly's Test of Sphericity and a two-way repeated measures ANOVA were performed for each dependent variable. If there was a significant interaction effect between the independent variables, then a Mauchley's Test of Sphericity and a one-way repeated measures ANOVA was performed for each independent variable. If Mauchly's Test of Sphericity was violated, we used a lower bound estimate for $F$ and $p$ values from ANOVA indicated by $F^{*}$ and $p^{*}$. We calculated the effect size for each component of the repeated measures ANOVA using Partial Eta Squared. If any independent variable or combinations had statistically significant effects ($p < 0.05$), Bonferroni-corrected post-hoc tests were used to determine which pairs were significantly different. 
During the study, there were minor device malfunctions for 4 participants. Therefore, we did not include those 4 participants and conducted all of our analyses with 20 participants.

\subsection{Results}
\begin{figure}[t]
\centering
\vspace{0.05in}
  \includegraphics[width=\columnwidth]{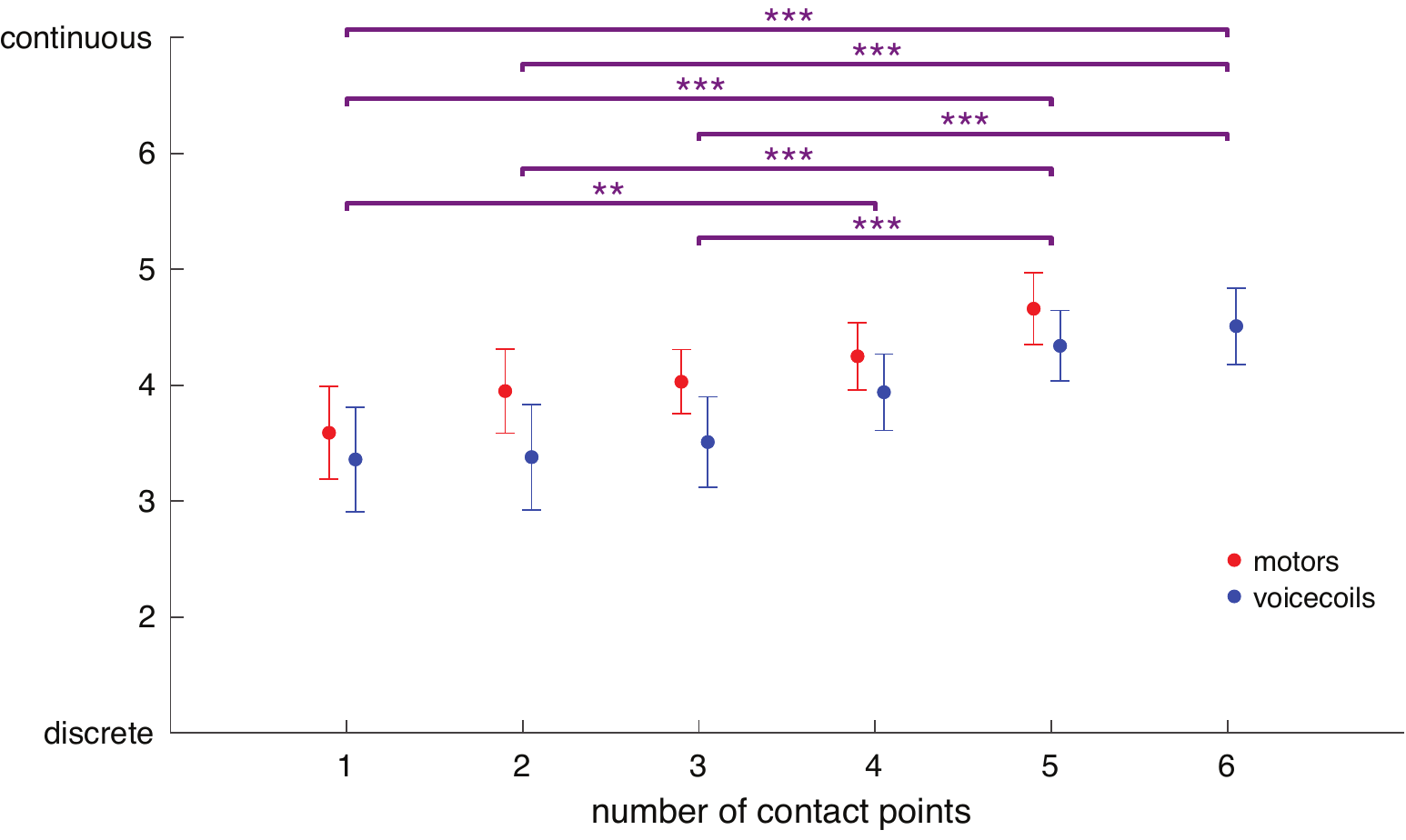}
  \caption{\newText{Average continuity ratings of all participants with standard error bars. Results of the two-way repeated measures ANOVA show that the devices are statistically significantly different from one another. Statistical significance from the two-way repeated measures ANOVA pertaining to number of contact points shown in purple.}}
  ~\label{fig:continuity}
\end{figure}

\begin{figure}
\centering
  \includegraphics[width=\columnwidth]{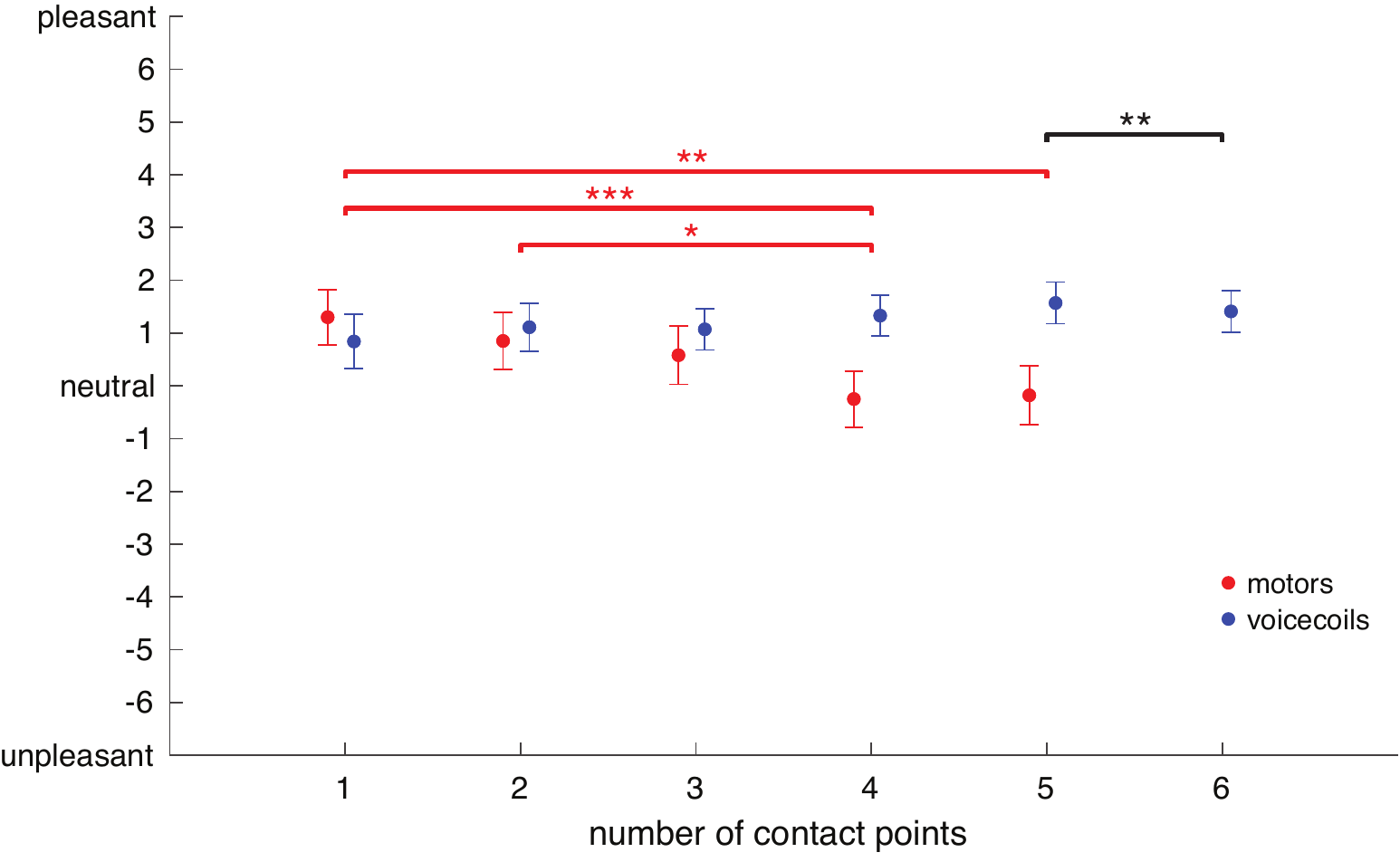}
  \caption{\newText{Average pleasantness ratings of all participants with standard error bars. Results of the two-way repeated measures ANOVA show that the devices are statistically significantly different from one another. Statistical significance from the result of the one-way repeated measures ANOVA of the motor device with 5 contact points and the voice coil device with 6 contact points shown in black.
  Statistical significance from the result of the one-way repeated measures ANOVA of the number of contact points for the motor device (performed due to the significant interaction effect between device and number of contact points from the two-way repeated measures ANOVA) shown in red.}}
  ~\label{fig:pleasantness}
\end{figure}

Figure~\ref{fig:continuity} and Figure~\ref{fig:pleasantness} report the means of all dependent variables for each haptic parameter along with their standard errors and significance ($*: 0.01 < p < 0.05$, $**: 0.001 < p < 0.01$, $***: p < 0.001$).

\subsubsection{Original Designs of the SHIFTS Devices}
The results of the one-way repeated measures ANOVA of the original designs of the SHIFTS devices (motor device with 5 contact points and voice coil device with 6 contact points) showed that there was not a significant difference in the continuity ratings between devices. However, there was a significant difference in the pleasantness ratings ($F^{*}(0.25,49.5)=18.98$, $p^{*}=0.006$, ${\eta_p}^{2} = .087$). This significant difference is shown in Figure~\ref{fig:pleasantness} in black.

\subsubsection{Actuation Type and Number of Contact Points}
We ran a two-way repeated measures ANOVA on the continuity ratings with SHIFTS device and number of contact points as factors. This analysis showed that there was a significant difference in continuity ratings between the devices ($F^{*}(0.25,247.5)=11.36$, $p^{*}=0.017$, ${\eta_p}^{2} = .011$)
\camready{ and} between the number of contact points ($F^{*}(1,247.5)=9.72$, $p^{*}=0.002$, ${\eta_p}^{2} = .038$), but there was no significant interaction between SHIFTS device and number of contact points ($F^{*}(1,247.5)=0.32$, $p^{*}=0.572$, ${\eta_p}^{2} = .001$). The results of the post-hoc test with a Bonferroni correction confirmed that there was a significant difference in continuity ratings between the SHIFTS devices ($p < 0.001$). The post-hoc test also showed that the ratings for only 1 contact point is significantly different from 4 ($p < 0.01$), 5 ($p < 0.001$), and 6 ($p < 0.001$) contact points, that 2 contact points are significantly different from 5 ($p < 0.001$) and 6 ($p < 0.001$) contact points, and 3 contact points are significantly different from 5 ($p < 0.001$) and 6 contact points ($p < 0.001$). These results are shown in purple in Figure~\ref{fig:continuity}.

We ran a two-way repeated measures ANOVA on the pleasantness ratings with SHIFTS device and number of contact points as factors. This analysis showed that there was a significant difference in \camready{pleasantness }
ratings between the devices ($F^{*}(0.25,247.5)=21.48$, $p^{*}=0.003$, ${\eta_p}^{2} = .021$), but not between the number of contact points ($F(1,247.5)=1.49$, $p=0.223$, ${\eta_p}^{2} = .006$). However, there was a significant interaction between SHIFTS device and number of contact points ($F^{*}(1,247.5)=7.08$, $p^{*}=0.008$, ${\eta_p}^{2} = .028$). The results of the post-hoc test with a Bonferroni correction confirmed that there was a significant difference in \camready{pleasantness }
ratings between the SHIFTS devices ($p<0.001$). 

Since there was a significant interaction between SHIFTS device and number of contact points, we ran a one-way repeated measures ANOVA for each SHIFTS device with number of contact points as the factor. The results of the one-way repeated measures ANOVA for the motor device showed a significant difference between the pleasantness ratings for different numbers of contact points ($F^{*}(1,123.75)=6.17$, $p^{*}=0.014$, ${\eta_p}^{2} = .047$). The post-hoc test with a Bonferroni correction showed that the ratings for only 1 contact point is significantly different from 4 ($p < 0.001$) and 5 ($p < 0.01$) contact points and that 2 contact points are significantly different from 4 ($p < 0.05$) contact points. These results are shown in red in Figure~\ref{fig:pleasantness}. The results of the one-way repeated measures ANOVA for the voice coil device showed that there was no significant difference between the pleasantness ratings for different numbers of contact points ($F^{*}(1.25,148.5)=1.42$, $p^{*}=0.241$, ${\eta_p}^{2} = .012$).

In the post-study survey, for the 20 participants that were included in the analysis, 12 stated that they preferred the voice coil device and 8 preferred the motor device. In the space available to describe differences in the sensations, 3 participants specifically stated that the voice coil device felt ``natural" 
and 6 participants stated that they felt ``smoother". 
Pertaining to the motor device, several participants insinuated, and 1 specifically stated, that the sensation felt more ``continuous". 
Another participant stated that the sensation felt ``like 
a human touching my arm". In the space available to provide general comments, nearly every participant stated that they noticed the ``length" 
of the trial varying, implying either the total time of the applied sensation or the distance 
the sensation traveled along the forearm. One participant specifically stated that they preferred the sensation close to the wrist, but did not like the sensation close to their elbow. Additionally, 1 participant stated that it was difficult to know what the difference between continuous and discrete is without examples of a maximum and minimum condition and another participant stated it was difficult to rate the pleasantness because none of the sensations felt unpleasant. Finally, 1 participant stated they thought the sensation would have felt better or nicer if they had been warm.

\subsection{Discussion}
\subsubsection{Comparison of Original Design of SHIFTS Devices}
From our initial analysis, we found that there was no difference in continuity rating between the devices when comparing their original designs. However, there was a significant difference in the pleasantness ratings. Therefore, we can conclude that although there was no difference in the performance of the devices in creating a continuous stroke, the voice coil device is able to create a more pleasant sensation than the motor device. This result matches the participants' responses from the post-study survey in that more participants preferred the voice coil device.

\subsubsection{Continuity of Tactile Stroking}
From the results shown in Figure~\ref{fig:continuity}, we can easily see the significant effects that the SHIFTS devices and the number of contact points have on user response. The continuity ratings for the motor device were statistically significantly greater than the continuity ratings of the voice coil device. This matches our original hypothesis that incorporating direct lateral motion via sequential discrete lateral skin-slip would improve the haptic illusion of a continuous stroking sensation compared to simple normal indentation. Additionally, our results support our hypothesis that decreasing the number of contact points will decrease the continuity of the stroking sensation. Finally, since there is no significant difference in the continuity ratings between 4, 5, and 6 contact points, we conclude that 4 contact points are necessary to create an effective tactile stroking sensation.

\subsubsection{Pleasantness of Tactile Stroking}
From the results shown in Figure~\ref{fig:pleasantness}, there was a significant difference in the pleasantness ratings between the SHIFTS devices used to apply the sensation. The voice coil device was rated as significantly more pleasant than the motor device. This quantitative data matched the qualitative data collected via the post-study survey. While there was no significant difference in the pleasantness ratings across the different numbers of contact points with the voice coil device, the pleasantness rating for the motor device decreased as the number of contact points increased. However, the ratings for the sensation do not venture into feeling unpleasant. Differences in the design of the devices are likely responsible for the difference in the pleasantness ratings, as opposed to the actuation technique. The voice coil device is a wearable sleeve made of an elastic material and the participants were able to rest their arm on the table in front of them, which was likely more comfortable for the user than the motor device, which is not a wearable device and consists of more rigid, less conforming materials.\camready{ In our previous studies~\cite{VCstudy,NunezToH2019}, we investigated the control parameters of the motors SHIFTS device at both the dorsal and volar forearm, but only the dorsal forearm for the voice coil SHIFTS device. While we could have provided the stimulus for the motor SHIFTS device to the dorsal forearm to match the voice coil SHIFTS device, we chose to provide the stimuli to the volar forearm because we wanted to investigate the best performing conditions. Since the pleasantness ratings for the motor SHIFTS device are less pleasant than those of the voice coil SHIFTS device, we can conclude that if we had presented the stimuli to dorsal forearm, the perceived pleasantness would still be less than the perceived pleasantness from the voice coil SHIFTS device because our previous study showed that the volar forearm was significantly more pleasant than the dorsal forearm~\cite{NunezToH2019}.}

\subsubsection{Necessary Design Parameters}
Based on our experimental results, a minimum number of 4 contact points is necessary to effectively create tactile stroking (when keeping the distance between contact points and actuation profile constant at the values previously investigated~\cite{VCstudy,NunezToH2019}). Although integrating sequential discrete lateral skin-slip into a wearable device is more difficult, it will create a more continuous sensation than simple sequential discrete normal indentation and was directly compared by one participant to a human touching the their arm. However, there are still important parameters that we believe need to be investigated to fully define the optimal parameters needed to replicate tactile stroking similar to what is relayed by humans during social touch interactions. The effect that the speed, delay between actuation of adjacent actuators, distance, and now number of contact points has on creating a continuous, pleasant stroking sensation has been explored, but the role that contact area or temperature have on the sensation has yet to be investigated and would be interesting future work.\camready{ Another area for future work would be to control force rather than indentation distance into the skin, which could equalize haptic sensation magnitude across users with different skin stiffness.}

\section{CONCLUSIONS}

\label{sec: conclusion}
In this paper, we discussed the use of \newText{social haptic illusions for tactile stroking }
to replicate stroking on the forearm. We presented two SHIFTS devices that use haptic illusions to create continuous and pleasant stroking sensations. We discussed the differences in device actuation and conducted a user study to directly compare device performance as it pertained to perceived continuity and pleasantness. In addition to comparing the devices using the parameters that correspond to the most continuous and pleasant sensation, we assessed the effect that reducing the number of contact points has on the sensation. We found that while the motors SHIFTS device creates a more continuous sensation, the sensations created by the voice coil SHIFTS device was perceived as more pleasant. Importantly, we also found that one can use as few as four contact points to create a continuous and pleasant stroking sensation, thereby reducing the overall size and power needed to drive a wearable haptic device. This paper aims to spur interest and aid haptic designers in the development of future wearable haptic devices for unobtrusive, private, \camready{and }effective social communication.





\section*{ACKNOWLEDGMENT}

The authors would like to thank Benjamin Pratt for his assistance in building the SHIFTS devices \newText{and our collaborators at Facebook, Inc. for their ideas towards the design of the SHIFTS devices}.

\bibliographystyle{IEEEtran}
\bibliography{IEEEabrv,HS2020abbrev}

\end{document}